\documentclass[conference]{IEEEtran}
\usepackage{graphicx}
\usepackage{caption}
\usepackage{subcaption}

\usepackage{cite}

\ifCLASSINFOpdf
\else
\fi
\begin{document}
%
\title{Dashbell: A Low-cost Smart Doorbell System for Home Use}

\author{\IEEEauthorblockN{Bradley Quadros, Ronit Kadam\\
Devendra Lavaniya, and Muhammad Mukhtar }
\IEEEauthorblockA{
Department of EECS,
University of California, Irvine\\
Irvine, CA92617, USA\\
Email: \{bquadros, kadamr,  dlavaniy, mohamme\}@uci.edu}
\and
\IEEEauthorblockN{Kartik Saxena, Wen Shen and Alfred Kobsa\\ }
\IEEEauthorblockA{Department of Informatics\\
University of California, Irvine\\
Irvine, CA92617, USA\\
\{saxenak, wen.shen, kobsa\}@uci.edu}
}


%


\maketitle

\begin{abstract}
Smart doorbells allow home owners to receive alerts when a visitor is at the door, see who the guest is, and communicate with the visitor from a smart device. They greatly improve people's life quality and contribute to the evolution of smart homes.  However, the commercial smart doorbells are quite expensive, usually cost more than 190 US dollars, which is a substantial impediment on the pervasiveness of smart doorbells. To solve this problem, we introduce the Dashbell-a budget smart doorbell system for home use. It connects a WiFi-enabled device, the Amazon Dash Button, to a network and enables the home owner to answer the bell triggered by the dash button using a smartphone. The Dashbell system also enables fast fault detection and diagnosis due to its distributed framework.
\end{abstract}


%
\IEEEpeerreviewmaketitle

\section{Introduction}
Doorbells have been playing an important role in protecting the security of modern homes since they were invented~\cite{robles2010review}\cite{mohammad2009electronic}.  A doorbell allows visitors to announce their presence and request entry into a building as well as enables the occupant to verity the identity of the guests to help prevent home robbery or invasion at a moment's notice\cite{robles2010review}. There are two types of doorbells depending on the requirement of wall wiring: the wired doorbells and the wireless doorbells~\cite{shah2010doorbell}\cite{mohammad2009electronic}. The former requires a wire to connect both the front door button and the back door button to a transformer, while the latter transfer the signal wirelessly using telephone technology~\cite{shah2010doorbell}\cite{mohammad2009electronic}.

Modern buildings are typically equipped with wireless doorbell systems that employ radio technology to signal doorbells and answer the doors remotely. Although these doorbells are much more convenient than wired ones, they do not always satisfy the demands of modern homes for the following three reasons. First, the answering machines are normally located at a fixed place (often near to the door), if a occupant wants to answer the doorbell, he/she has to go to the answering machines. Second, if the occupant would like to see the visitors outside, he/she has to go to door. Third, the occupant has no way to answer or admit guests when he/she is not at home, nor to keep a record of guests.

As smart home technology matures, smart doorbells can solve this problem greatly by connecting the doorbells to the Internet(or a local network) and allowing users to answer the bell through a smart device such as a smartphone or tablet~\cite{senagala2006rethinking}\cite{augusto2006smart}\cite{cook2004smart}\cite{vig2010smart}. This enables a home owner to answer and admit a visitor anywhere when a smart device connecting to the Internet is available~\cite{demiris2008technologies}\cite{han2010design}. However, such smart doorbells are quite expensive (usually more than 190 US dollars) due to technical and manufacturing difficulties~\cite{higginbotham2015connected}\cite{leclair2014smart}\cite{baker2015seewhen}. The high prices make these products unavailable to most home users with limited budgets, hindering the pervasiveness of smart doorbells. This is confirmed by a research shows that less than $4\%$ of U.S. households have a smart doorbell system to protect the security of the homes~\cite{maia2014evolution}. To solve this problem, we introduce the Dashbell- a low-cost smart doorbell system for home use. The doorbell system uses a cheap, WiFi-enabled device-the Amazon Dash Button to serve as the doorbell, and connects it to the Internet, allowing users to answer the bell anywhere using a smart device such as a smartphone or a tablet. With such as solution, users may purchase a smart doorbell system at a price as low as 40 US dollars, which significantly increases the affordability of the smart doorbells.

\section{Related Work}
\subsection{Smart Doorbells}
A smart doorbell is an integral part of a smart home, which helps protect the security of the home by avoiding unwanted access such as robbery and invasion~\cite{senagala2006rethinking}\cite{augusto2006smart}\cite{cook2004smart}\cite{vig2010smart}. The controller of the smart home can potentially answer the bell and decide whether to admit a visitor outside the door or not through adaptive learning and other technologies~\cite{lehmann2004home}. Because of the important role that smart doorbells play on building a smart home, many techniques and methodologies have been invented during past few years\cite{vig2010smart}\cite{huisking2012method}\cite{yang2015smart}\cite{fadell2014visitor}.  

The existing smart doorbells provide an integrated solution, which means that the working mechanisms or the implementation details are hidden and unknown to the users. If there is a failure, users have to seek help from professionals for repairs or maintenance. It is also very likely that users need to replace the whole smart doorbells  due to a failure of a component in the system.

\subsection{Dash Button}
Amazon Dash Button is a WiFi-enabled device that allows consumers to reorder frequently used daily products like trash bags, toilet towels or refill blades~\cite{amazon2015dash}\cite{fernandez2015thinking}\cite{ellis88honoring}\cite{crouch2015horror} by pressing a button. A dash button can be purchased via online and costs 4.99 US dollars~\cite{amazon2015dash}.  Recently, it has been found that the dash button can be tweaked to track baby habits~\cite{benson2015hack}. We employ this feature of the dash button and use it as a doorbell (or a trigger) of the Dashbell system.

Alternatives to Dash Button include portable door bell kits,  and wireless door chime trigger with motion sensors, both of which can be purchased for less than $5$ US dollars.


\section{Dashbell: A Low-Cost Solution}

\begin{figure*}[ht!]
\centering
\includegraphics[width=.95\linewidth]{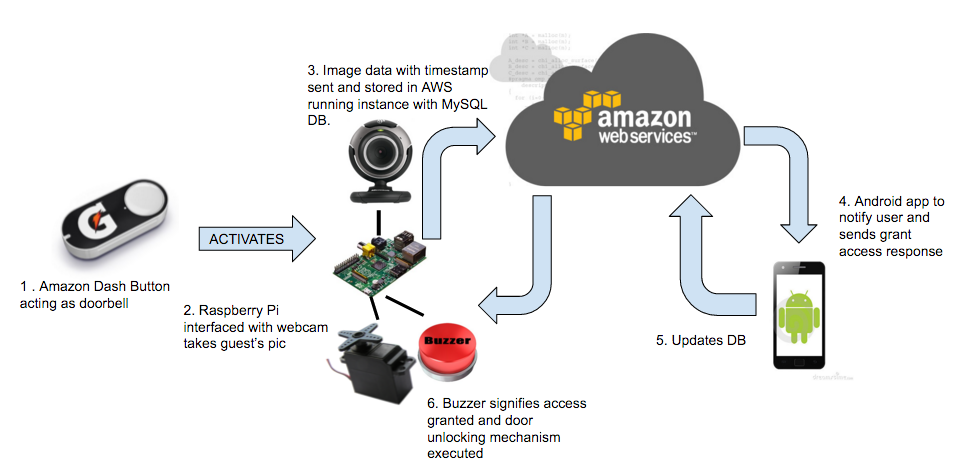}
\caption{The architecture of DashBell.}
\label{fig:framework}
\end{figure*}

The Dashbell enables the home owner to see the pictures of a visitor to remotely ascertain their identity.  Besides, it allows the home owner to admit or reject a request anywhere when connected to the Internet. The Dashbell system (as shown in Fig.~\ref{fig:framework}) consists of a dash button(as the doorbell), a WiFi router, a computing device(e.g.,  a Raspberry Pi), a webcam, a buzzer, cloud computing service (e.g., Amazon Web Service), a smartphone and the Internet. The working flow  is described as follows:
\begin{itemize}
\item[1)]A visitor presses the dash button to request for entry to the home. 
\item[2)] A signal from the Dash button to the WiFi router is captured by the Raspberry Pi. It triggers the following operations: activate the webcam and take a picture of the guest who pressed the button; activate the buzzer for the alarm sound so that the owner is notified on the presence if inside the house; the image captured along with the timestamp is uploaded to the cloud.
\item [3)]The webcam takes pictures of the guest and process and sends them to the cloud database service (We use Amazon Web services).
\item [4)]The service notifies the user through the mobile app (We develop an Android app). 
\item[5)]The home owner decides whether to grant access to the visitor. The decision is updated on the server, which also delivers this message to the Raspberry Pi. If the home owner grants the access to the visitor, the door automatically opens using a servo-motor. If the access is denied, no actions will be taken and the guest's request for entry is denied.
\end{itemize}

\begin{figure*}[ht!]
\centering
\begin{subfigure}{.32\textwidth}
\label{fig:uia}
 \centering
\includegraphics[width=.95\linewidth]{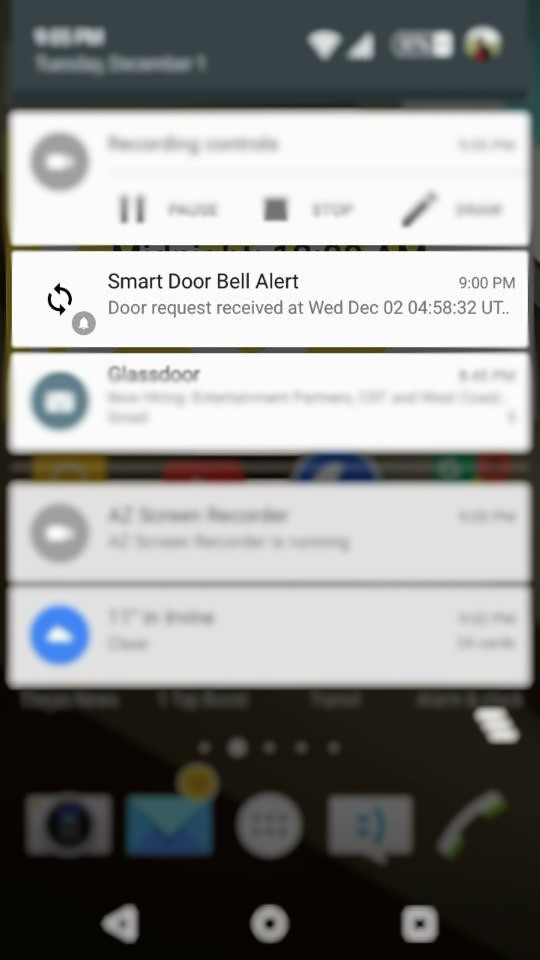}
\caption{Notification}
\end{subfigure}
\begin{subfigure}{.32\textwidth}
\label{fig:uib}
 \centering
\includegraphics[width=.95\linewidth]{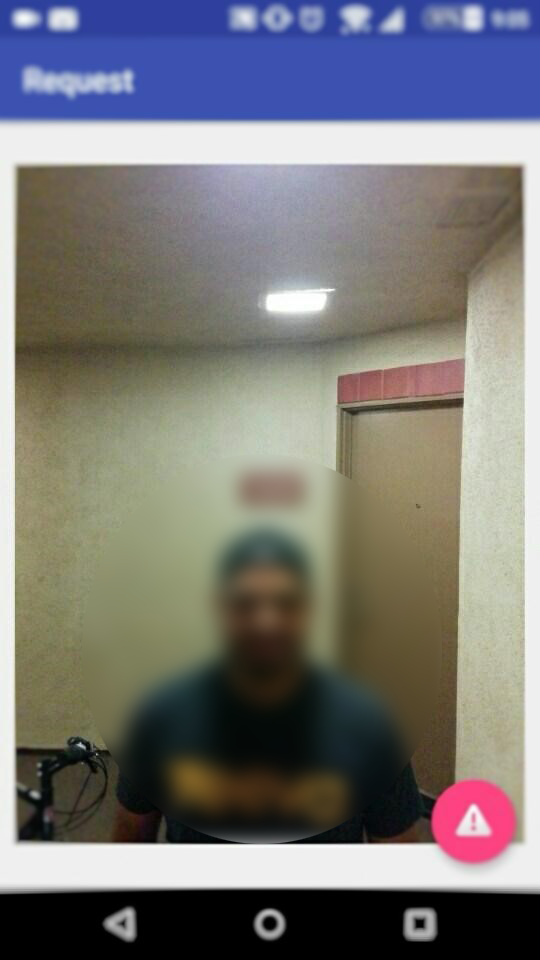}
\caption{Visitor Picture}
\end{subfigure}
\begin{subfigure}{.32\textwidth}
\label{fig:uic}
 \centering
\includegraphics[width=.95\linewidth]{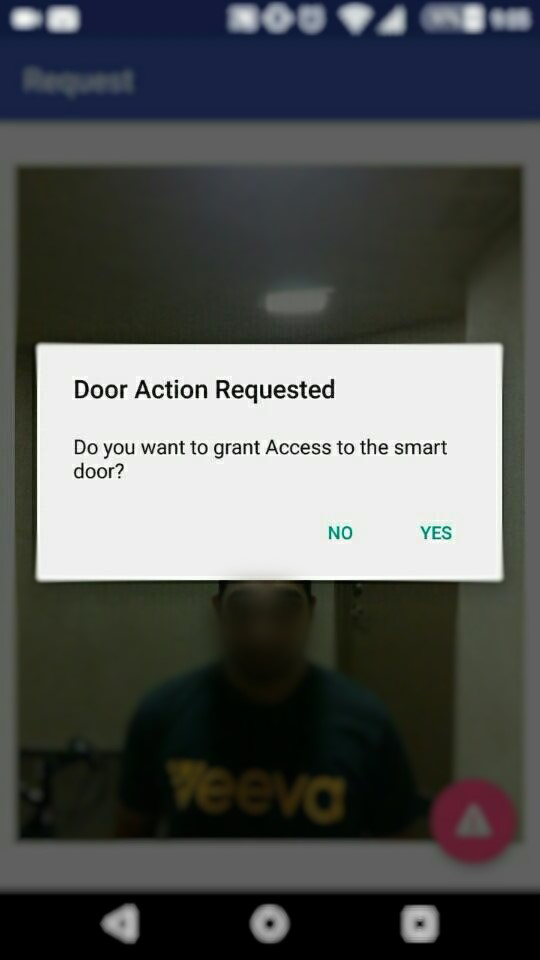}
\caption{Grant Access}
\end{subfigure}
\caption{The user interface of the mobile application:  (a) shows the notification on the phone received from the server.(b) Shows the picture of the visitor on the mobile application. The owner can click on the alert icon in the bottom right corner to show the access prompt. (c) shows the screen where the use can choose to accept or decline the visitor which will subsequently the door based on their decision. We will also have the option to scroll across a timeline and expand on the request made on a particular day. On clicking a particular request the owner will be directed to the next screen at where he / she will have a more detailed result of the request made at the selected time. The owner will also be provided with the settings page as shown in where configurable options for service enabling and notifications will be mentioned.}
\label{fig:ui}
\end{figure*}

The home owner has a variety of options to choose regarding this call: accept the call or reject the call if the service is on, or ignore the call if the "Do Not Disturb" is on. If the user grants the access of the guest, the message will be delivered to the Raspberry Pi by updating the database.  The owner is able to view the historical record of the requests. The user interface also enables the user to choose three types of alerts: Email, Text Message, and Ringer. The user interface of the mobile application is shown in Fig.~\ref{fig:ui}. 

We are using Amazon Web Service (AWS) cloud services for this implementation. We have a EC2 linux instance running which hosts a MySQL DB server as well as an Apache Web Server. The server maintains a TCP connection with the Raspberry Pi by listening on a specific port. When it receives the data object with the timestamp and the picture, it will be uploaded to the web server and a link is shared with the DB server, which maintains a database for all the entries. Additionally it maintains a column for ``Access Granted" as a binary``Yes" and ``No" value, initialized as ``Null" and updated with the user's response . Another TCP connection is maintained by the server with the user's mobile client on a separate port. It pushes when new data is received to the user's mobile device as a notification triggered by the mobile application.
When the user clicks on the notification, it shows the picture of the user and allows user to either grant or deny access as per his/her discretion. Other features offered by the mobile applications are explained below. The response is relayed back to the server and ``Access Granted" field is updates as ``Yes" or ``No" based on the response. This triggers another action in DB server and the response data is sent back to the Raspberry Pi.


\section{Discussion}
Modern homes are typically equipped with Wi-Fi routers and have access to the Internet. Smartphones  are also highly available to the majority of  population. To build a budget smart doorbell system like Dashbell, the user only needs to purchase a Amazon Dash Button (5 dollars), a Raspberry Pi (25 dollars),  a webcam (8 dollars), and a buzzer(1 dollar). The user can sign up to request for free Amazon Web Service. The total cost of a Dashbell system is less than 40 US dollars. 

The Dashbell system differentiates existing smart doorbell systems in the following aspects. First, Dashbell is much cheaper than existing smart doorbells. Second, Dashbell is a distributed system rather than an integrated one, which enables faster fault detection and diagnosis. For instance, if some of the components fail to operate, one can just identify and fix or replace the parts by checking each individual device instead of disassembling or replacing the whole doorbell system. Third, given that most smart doorbell devices are expensive devices, they can be potentially detached and stolen. However, with the Dashbell, only the Dash button, which is inexpensive and replaceable, needs to be placed outside the home, making it a much better alternative in terms of the device's own security. Lastly, unlike exisiting smart doorbells, which are only sold in limited places and through particular channels, the components of the Dashbell system are highly available.

While the Dashbell provides several useful features and enhanced security over a conventional doorbell, there are a few security and privacy issues associated with it. Since the device is connected via a home WiFi network, it is possible to compromise the network and use the device, grant access to unauthorized visitors or collect data using it without the owners consent. We advise users to keep their network secured with a password. Our system also takes pictures of visitors without their consent and stores them on the server. Since this is personally identifiable information, we have made sure the server and all communication channels are secure. Using secure communication channels and encrypting user data while storing on the server would be helpful in this regard. The mobile application also has an additional layer of security so that only owner can grant access to a visitor.

\section{Future Work}
In future work, we plan to implement one-way or two-way audio and video communication between the visitor and the homeowner. Further, we will integrate face recognition, natural language processing and machine learning techniques (e.g., Support Vector Machines, Neural Networks) to allow the Dashbell system to answer the door automatically without human intervention. 

We found that the internals of the button are much more interesting and powerful than advertised~\cite{baguley2015hack} as it contains a proprietary chip inside which contains a small ARM processor, flash memory and a microphone besides the WiFi module. However, due to inaccessible SDK and no APIs, it is nontrivial to use them. Nevertheless, if similar programmable devices with open APIs can be used, it is possible to create even lower cost doorbells with the button as a standalone device and microphone to record guests' voice and identity to make decisions.

We also propose to conduct a user study to evaluate the reliability and the usability of the Dashbell system upon finishing the development. This will also help us understand if users have any privacy concerns regarding the use of this device.

\section{Conclusion}
To increase the availability of smart doorbells, we introduce the Dashbell-a  smart doorbell system that utilizes the Amazon dash button and existing devices in modern homes to develop a budget solution for users. The Dashbell system has the similar functionality with its commercialized counterparts with a much lower price. It costs less than 40 US dollars compared to 199 US dollars for a single smart doorbell available in market. The Dashbell employs a distributed approach, which enables fast fault detection and diagnosis.  By doing so, we spread the benefits that smart doorbells bring to home users and contribute to building a smarter and better world through technology.






%

\bibliographystyle{IEEEtran}
\bibliography{percom}

\end{document}